# Decidability of Existence and Construction of a Complement of a given function

Ka.Shrinivaasan, Chennai Mathematical Institute (CMI)
(shrinivas@cmi.ac.in)

April 28, 2011

### Abstract

This article defines a complement of a function and conditions for existence of such a complement function and presents few algorithms to construct a complement.

## 1  Goal

Goal is to analyze decidability of finding a complement of a function defined over both domain and range as integers. There are two cases to consider:

1. The domain and range of the function are infinite
2. The domain and range of the function are finite

## 2  Example of a complement function (when the domain and range are infinite)

Consider an infinite set of positive integers = $\{1, 2, 3, 4, 5...\}$ . Let us say we create a subset of even integers out of it = $\{2, 4, 6, 8, ...\}$. This set of even numbers can be generated by using the function $f(x) = 2x$, x=0,1,2,3... A complement of this set is infinite set of odd numbers and they can be generated using g(x) = 2x+1 , x=0,1,2,3,4,... So we define $g(x) = 2x + 1$ to be *a complement function* of $f(x) = 2x$ over the set of integers.
Similarly for the function $f(x, y) = xy$, a complement function (ignoring trivial factors 1 and n) is over the set of all primes. Thus finding a complement g of f(x,y) amounts to finding distribution of primes i.e g shows a pattern in primes.This was a question posted in google group sci.math (*http* : *//groups.google.com/group/sci.math/browse*_*thread/thread* */46ab0335ce9205d9/f3c67ee19926e02e*)

# 3 Existence of a complement when domain and range are finite

## 3.1 Definitions

1. Let U be a universe of n-bit integers of size $2^n$ integers

2. Function $f_n$ is defined over the domain of a-bit integers and range U of n bit integers - f:$\{0,1\}^a$ - $\{0,1\}^n$ for $a > 0$ and $n > 0$

3. Function $g_n$ is defined over the domain of b-bit integers and range U of n bit integers - g:$\{0,1\}^b$-$\{0,1\}^n$ for $b > 0$ and $n > 0$

4. $L(f_n)$ = language generated by the function $f_n = \{f(x_0), f(x_1), ..., f(x_{2^a-1})\}$

5. $L(g_n)$ = language generated by a complement function $g_n = \{g(x_0), g(x_1), ..., g(x_{2^b-1})\}$

6. Function $g_n$ is defined as *a complement* of $f_n$ if

    (a) $g_n$ is total
    
    (b) $g_n(x) \neq f_n(x) \quad \forall x$
    
    (c) $L(f_n) \cup L(g_n) = \{0,1\}^n$ and
    
    (d) $L(f_n) \cap L(g_n) = \phi$

7. The subscript n for the functions f and g denotes the finiteness of the size of input and output.

## 3.2 Conditions for existence of a complement

For a complement to exist the universe U - the range of functions $f_n$ and $g_n$ - must be partitioned by the functions $f_n$ and $g_n$. Due to this, if

1. $f_n$:$\{0,1\}^a$ - $\{0,1\}^n$ and

2. $g_n$:$\{0,1\}^b$ - $\{0,1\}^n$

then the inequality $2^a + 2^b >= 2^n$ must hold. This is because, function $f_n$ maps $2^a$ elements to X number of elements ($X <= 2^a$) in U and function $g_n$ maps $2^b$ elements to Y elements ($Y <= 2^b$) in U. This makes $X + Y <= 2^a + 2^b$. Since $X + Y = 2^n$ by definition of complement, $2^a + 2^b >= 2^n$.

## 3.3 Algorithm for complement construction

Input to a complement construction algorithm is the function $f_n$ and the output is a complement function $g_n$ of $f_n$.(Univariate function (polynomial) has been assumed. This can be generalized to multivariate polynomials.). Algorithm comprises of 3 high-level steps.

### 3.3.1 STEP 1-Getting the range set for complement function

From the function $f_n$ we can get the set T = $\{f_n(i)/0 \leq i \leq 2^a - 1\}$.From this the set S = $U \setminus T$ can be constructed.

### 3.3.2 STEP 2-Construction of mapping

Before we construct a function representation, a mapping has to be established between a set $P = \{0, 1, 2, 3, ..., 2^b - 1\}$ and the set S. This mapping is a lookup table which imposes an ordering on the set S. We have that $|S| = 2^n - |T|$. For a valid function to exist, we must have $2^b \geq 2^n - |T|$ (since for a valid function, domain must be greater than or equal in size to range) and hence we have to choose b such that $b \geq log_2(2^n - |T|)$. Number of possible mappings(and hence functions) between P and S is $(|S|)^{|P|}$. But this includes mappings which are not onto also. To cover set S, we have to choose one onto mapping to extract a complement function out of various possible complement functions. One such mapping is done as follows

1. i=0

2. while($i <= 2^b$)

    (a) If set S is not empty, choose uniformly at random, an element $e$ from set S and map it to i and add to the lookup table as $< Key : Value >$ pair $< i : e >$. Remove element $e$ from the set S. This implies $g_n(i) = e$ where $g_n$ is a complement to be constructed.

    (b) else if the set S is empty, map the first element in the ordering $g_n(0)$ to i by adding $< i : g_n(0 >$ to lookup table i.e $g_n(i) = g_n(0)$.

    (c) i := i+1

The above mapping procedure selects one mapping out of $|S|!$ onto mappings (since after adding each element to lookup table, number of possible candidates reduces by 1 element from previous iteration). Many other ways of mapping are possible different from the procedure above. Each one of these mapping procedures result in different complement function.

### 3.3.3 STEP 3-Construction of complement function representation from mapping

Once we have the mapping constructed as in previous step, we can either apply 1) polynomial interpolation (or) 2) Arithmetization using simple replacement of boolean expressions with arithmetic expressions (or) by fourier polynomials 3) or by applying lambda calculus to get the function representation for a complement. These 3 algorithms for construction of complement function from mapping are explained in detail next.

<u>Algorithm   1 - By polynomial interpolation</u>

1. Since we have been given with a mapping from previous step between $x_i \epsilon \{0, 1\}^b$, and $y_i \epsilon S$, we can apply polynomial interpolation theorem which states that given $x_1, y_1, x_2, y_2, ..., x_n, y_n$ with $x_i \neq x_k$ for $i \neq k$, $\exists$ polynomial $p^{(n)} \epsilon F[x]$ of degree less than or equal to n-1, such that $p^{(n)}(x_i) = y_i$. The recurrence for building the polynomial is as follows:
$p^{(i+1)}(x) = p^{(i)}(x) + \{[y_{j+1} - p^{(i)}(x_{j+1})]/q^{(i+1)}(x_{j+1})\} * [q^{(i+1)}(x)]$
where $q^{(i+1)}(x) = \prod_{j=1}^{i}(x - x_j)$

<u>Algorithm   2 - By representing boolean function by polynomial:</u>

1. From the mapping obtained in previous section, we can construct function representation for a complement. Let binary of $x = s_b s_{b-1}...s_3 s_2 s_1$ and binary of $Mapping(g_n)(x) = z_n z_{n-1}...z_3 z_2 z_1$ obtained by looking up x from mapping lookup table constructed earlier. The $i^{th}$ bit of $Mapping(g_n)(x)$ is denoted as $g_n^i(x)$.

2. Construct a DNF boolean function for $z_i$ with $2^b$ clauses where each clause corresponds to a binary string in $\{0,1\}^b$ with binary digits replaced by variables as: $z_i = (\neg s_b \cap \neg s_{b-1} \cap ... \cap \neg s_3 \cap \neg s_2 \cap \neg s_1 \cap g_n^i(x_0)) \cup (\neg s_b \cap \neg s_{b-1} \cap ... \cap \neg s_3 \cap s_2 \cap \neg s_1 \cap g_n^i(x_1)) \cup ...(s_b \cap s_{b-1} \cap ... \cap s_3 \cap s_2 \cap s_1 \cap g_n^i(x_{2^b-1}))$. For example clause for a binary string 001010 with r=6 will be $(\neg s_6 \cap \neg s_5 \cap s_4 \cap \neg s_3 \cap s_2 \cap \neg s_1)$. The bits of $Mapping(g_n)(x)$ will be hard-coded in the formula above from mapping already constructed.

3. The boolean function above outputs the $i^{th}$ bit of $g_n(x)$ where $binary(x) = s_b s_{b-1}...s_3 s_2 s_1$. Intuitively, this boolean function is a multiplexer which selects the i-th bit for $g_n(x)$ among $2^b$ candidate bits. We have to construct n boolean functions as above for each of the n bits of $g_n(x)$

4. These boolean functions can be minimized to obtain minimum equivalent expression. After minimization we can arithmetize the boolean functions in one of the two ways below to get arithmetic expressions(polynomials):

    (a) <u>Simple Arithmetization</u>
       i. In the above boolean expression, replace a positive literal $s_i$ by a variable $s_i$ and a negative literal $(\neg s_i)$ by an arithmetic expression $(1 - s_i)$.
       ii. In the above boolean expression, replace a subexpression of the form $y_m \cap y_n$ by the arithmetic expression $t_m * t_n$
       iii. In the above boolean expression, replace a subexpression of the form $y_m \cup y_n$ by the arithmetic expression $1 - ((1 - t_m) * (1 - t_n))$
       iv. Repeat above steps recursively till the full boolean function for $z_i$ gets arithmetized.
       v. At the end of above algorithm the whole boolean function for $z_i$ is arithmetized as a polynomial in variables $s_b, s_{b-1}, ..., s_3, s_2, s_1$ which are the bit positions of input x. Let us denote this by $Arith(z_i)$. Since we have arithmetic expressions for individual bit positions of $g_n(x)$ in terms of bit positions of x, we can construct $g_n(x)$ by weighted summation given as $g_n(x) = 2^n * Arith(z_n) + 2^{n-1} * Arith(z_{n-1}) + ... + 1 * Arith(z_1)$. This is the required complement function $g_n(s_b, s_{b-1}, ..., s_3, s_2, s_1)$

    (b) <u>Arithmetization by fourier expansion</u>
       i. Since we have a boolean function for each bit position $z_i$ in step 2, the RHS of each $z_i$ can be expanded as a fourier polynomial in variables $s_b, s_{b-1}, ..., s_3, s_2, s_1$ which are the bit positions of input x. Thus we have a polynomial for i-th bit of $g_n(x)$ in terms of bit positions of x. Let us denote this expansion as $fourier(z_i)$
       ii. Thus $g_n(x)$ can be represented as the weighted summation of the bit positions of $g_n(x)$ as : $g_n(x) = 2^n * ((fourier(z_n) + 1)/2) +$

$2^{n-1}*((fourier(z_{n-1})+1)/2)+...+1*((fourier(z_1)+1)/2)$ since fourier polynomials evaluate to $\{-1,1\}$ and this transformation is needed to get the bits $\{0,1\}$

   iii. Complement function $g_n(s_b, s_{b-1}, ...s_3, s_2, s_1)$ thus has been constructed by applying fourier representation ($g_n(x)$ is a polynomial in the bit positions of x ($s_b, s_{b-1}, ...s_3, s_2, s_1$ ))

Algorithm  3 - applying Lambda Calculus:

1. Obtaining $g_n(x)$ given x, suffices for showing the existence complement. First we develop an iterative procedure for returning $g_n(x)$ if x is given as follows:

2. Procedure GetGofX(x):

   (a) u:=0
   (b) z:=-1
   (c) choose b such that $b \geq log_2(2^n - |T|)$ where T = $\{f_n(i)/0 \leq i \leq 2^a - 1\}$.
   (d) if ( $x > 2^b - 1$) printerror "error: out of range"
   (e) while($u \leq 2^n - 1$ and $z \leq 2^b - 1$)
       i. if u is in T then u:=u+1
       ii. if u is not in T then
           A. z:=z+1
           B. if($u \leq 2^n - 1$)
           C. {
           D. if(z equals x) add the entry $< x : u >$ to mapping table to signify $g_n(x) = u$ and return u as $g_n(x)$
           E. else add the entry $< z : u >$ to mapping table to signify $g_n(z) = u$ and set u:=u+1
           F. }
           G. else break
   (f) if($u \geq 2^n$) then add the entry $< x : g_n(0) >$ to mapping table to signify $g_n(x) = g_n(0)$ and return $g_n(0)$ as $g_n(x)$ by looking up mapping table for key 0

3. Correctness of the above procedure can be proved since we are guaranteed that $2^b \geq (2^n - |T|)$ by choice of b. The while loop breaks when all the $2^n$ elements fall either into $L(f_n)$ or to $L(g_n)$ and similar to previous section on construction of a mapping, we hardcode $g_n(x) = g_n(0)$

4. The inequality $2^a + 2^b \geq 2^n$ must hold as mentioned earlier in section on conditions for existence of complement (where a and b are input sizes to f and g respectively).

5. Now that we have an iterative procedure for $g_n(x)$, we can convert this into lambda expression by replacing loops with a fixed point combinator from lambda calculus. The equals() relation above is bitwise equivalence represented by literals for each bit. This lambda expression can again be converted to a standard logical formula(for example a boolean formula) in terms of bit positions.

# 4 Existence of a complement when domain and range are infinite

## 4.1 Definitions

1. Function $f$ is defined over the domain of integers and range of integers - f:$Z - Z$

2. Function $g$ is defined over the domain of integers and range of integers - g:$Z - Z$

3. $L(f)$ = language generated by the function $f = \{f(x_0), f(x_1), ...\}$ - an infinite set

4. $L(g)$ = language generated by a complement function $g = \{g(x_0), g(x_1), ...\}$ - an infinite set

5. Function $g$ is defined as *a complement* of $f$ if

    (a) $g$ is total
    (b) $g(x) \neq f(x) \quad \forall x$
    (c) $L(f) \cup L(g) = \{1, 2, 3, 4, 5, ...\}$ and
    (d) $L(f) \cap L(g) = \phi$

6. The subscript n for the functions f and g has been removed compared to previous section.

## 4.2 Decidability of complementation

To find an algorithm which would give a complement function g, given input function f. There are two cases to consider:

1. $L(f)$ is not recursive but recursively enumerable.
2. $L(f)$ is recursive

## 4.3 Case 1 - $L(f)$ is not recursive but recursively enumerable

Q - $L(f)$ recognizer

1. for all x

    (a) if $y == f(x)$ then return "y is in $L(f)$"

2. return "y is not in $L(f)$"

$L(f)$ is recursively enumerable since TM Q, outputs yes if a string y is in $L(f)$ ($y = f(x)$ for some x) but loops when y is not in $L(f)$. But $L(f)$ is not recursive since TM Q does not halt on all inputs. From definitions above we have complement of $L(f) = L(g)$. [Question of if Q outputs yes/no for any y is equivalent to asking if Q halts on any input. Reduction is defined as :

$x \; \epsilon \; L(Q) \iff R(x) \; \epsilon \; HaltingProblem$.] $L(f)$ is recursive if and only if both $L(f)$ and its complement $L(g)$ are recursively enumerable as proved below.

Claim: $L(f)$ is recursive if and only if both $L(f)$ and its complement $L(g)$ are recursively enumerable.

1. ($\Rightarrow$) If $L(f)$ is recursive then $L(f)$ is recursively enumerable. If $L(f)$ is recursive then complement of $L(f) = L(g)$ is recursive and hence $L(g)$ is recursively enumerable.

2. ($\Leftarrow$) Run two TMs M(f) and M(g) for $L(f)$ and $L(g)$. Input y. If y is in $L(f)$ then M(f) returns yes and M(g) loops. If y is not in $L(f)$ then M(f) loops and M(g) returns yes. Run M(h) accepting y which simulates M(f) and M(g). If M(f) return yes then M(h) returns yes and if M(g) returns yes then M(h) returns no. Thus M(h) returns yes or no for input y without looping.

Since $L(f)$ is not recursive, $L(g)$ is not recursively enumerable. So in this case, $L(g)$ has no turing machine that recognizes it.

## 4.4 Case 2 - $L(f)$ is recursive

$L(f)$ being recursive implies that there exists a turing machine which given input y always halts with accept($y \; \epsilon \; L(f)$)) or reject ($else$) states.

If $L(f)$ is recursive , $L(g)$ is also recursive by closure under complementation. This implies $L(g)$ has a turing machine deciding membership.

## 4.5 Decidability of construction of a complement function for infinite recursive set $L(g)$ - for Case 2 above

In previous section, $L(f_n)$ and $L(g_n)$ were finite languages. This restriction was helpful for defining finite DNF formulas for complement $g_n$ (or polynomial interpolation which is also for finite set of points). But for infinite recursive set $L(g)$, algorithms described in previous sections like polynomial interpolation or arithmetization do not work. This is because, we (either) might have to construct infinite boolean formulas and valuate such an infinite formula (or) might have to interpolate over infinite set of points. This procedure might never terminate. This can be proved undecidable by reduction from halting problem. Thus infinite cardinality for $L(g)$ makes construction of complement for infinite recursive sets undecidable.

## 4.6 Complementation as a nontrivial property - Application of Rice's theorem

1. Let Turing machine $P_f$ which accepts an integer x as input and writes $f(x)$ in output and goes to accept state. $P_f$ rejects if x is not an integer. $L(P_f) = L(f_n)$.. (A function can be thought of as a lambda expression which is equivalent to turing machine)

2. Let Turing machine $Q_g$ which accepts an integer x as input and writes $g(x)$ in output and goes to accept state. $Q_g$ rejects if x is not an integer.

$L(Q_g) = L(g)$.

3. Let Turing machine $X_i$ which accepts a pair of encodings of Turing machines $(P_f, Q_g)$. $L(X_i) = \{(P_f^k, Q_g^k)\}$, $k = 1, 2, 3, ....$ $X_i$ accepts only if input is an ordered pair of turing machine encoding of 2 functions over integers, and rejects otherwise.

4. Let Turing machine Y which accepts encodings of Turing machines $X_i$ and simulates $X_i$.
$L(Y) = \{X_i\}$, $i = 1, 2, 3, ...$

We want to test a property P such that language decided by $X_i$ - the set of ordered pairs of TM encodings - are complements of each other. This property P is nontrivial since not all ordered pairs are in P and if $L(X_m) = L(X_n)$ then both $X_m$ and $X_n$ are in P. Since by Rice's theorem it is undecidable if a language decided by a Turing machine has a nontrivial property, L(Y) is undecidable.

# 5 References

1. Lecture Notes on Algebra and Computation, MadhuSudhan, MIT (for Polynomial Interpolation)

2. Various literature on Lambda Calculus on internet

3. A question posted to Google Group sci.math $http://groups.google.com/group/$-

    (a) $/sci.math/browse\_thread/thread/46ab0335ce9205d9/f3c67ee19926e02e$

4. Complexity-I and Complexity-II Lecture Notes(2009 through 2010) - IMSc

5. Various literature on Rice's Theorem on internet

6. Automata Theory and Formal Languages - Aho, Hopcroft and Ullman

7. Introduction to Theory of Computation - Michael Sipser